\documentclass[journal, final]{IEEEtran}
\usepackage[lined, linesnumbered, ruled]{algorithm2e}
\usepackage{subfigure}
\usepackage{graphicx}
\usepackage{xcolor}

\usepackage{amsmath, amsthm, amssymb}

\begin{document}
\makeatletter
\@addtoreset{footnote}{page}
\renewcommand{\thefootnote}{\ifcase\value{footnote}\or 1 \or 2 \or * \fi}
\makeatother
\title{Modeling the Throughput of the Linux-based Agile-SD Transmission Control Protocol}
\author{Mohamed A. Alrshah$^1$,~\IEEEmembership{(Member,~IEEE),}
        Mohamed Othman$^{1*}$,~\IEEEmembership{(Member,~IEEE),}\\
        Borhanuddin Ali$^2$,~\IEEEmembership{(Member,~IEEE),}
        Zurina Mohd Hanapi$^1$,~\IEEEmembership{(Member,~IEEE)}
}

\markboth{IEEE Access, 2016}
{Alrshah \MakeLowercase{\textit{et al.}}: Modeling the Throughput of the Linux-based Agile-SD Transmission Control Protocol}

\maketitle

\footnotetext[1]{Department of Communication Technology and Network, Universiti Putra Malaysia, 43400 UPM, Serdang, Selangor D.E., Malaysia.}
\footnotetext[2]{Department of Computer and Communication Systems Engineering, Universiti Putra Malaysia, 43400 UPM, Serdang, Selangor D.E, Malaysia.}
\footnotetext[3]{The author is an associate researcher at the Computational \mbox{Science} and Mathematical Physics Lab, Institute of Mathematical Science, Universiti Putra Malaysia.\\
Corresponding authors: 
	\\ mohamed.asnd@gmail.com (Mohamed Alrshah),\\mothman@upm.edu.my (Mohamed Othman).}

\begin{abstract}
Agile-SD is one of the latest versions of loss-based Congestion Control Algorithm (CCA), which has been proposed to improve the total performance of TCP over high-speed and short-distance networks. It has introduced a new mechanism, called Agility Factor Mechanism (AFM), which shortens the epoch time to reduce the sensitivity to packet losses and in turn to increase the average throughput. Agile-SD has only been tested via simulation, however, it has not been mathematically proven or evaluated. The contribution of this paper is twofold: First, a new mathematical model for the throughput of NewReno and Agile-SD is proposed. This model is designed using the well-known Markov chains to validate the correctness of Agile-SD and to show the impact of buffer size, multiplicative decrease factor and maximum limit of agility factor ($\lambda_{\rm max}$) on the total performance. Second, an Automated Algorithm Configuration and Parameter Tuning (AACPT) technique is employed to optimize and automate the configuration of $\lambda_{\rm max}$. Further, the numerical results for both NewReno and Agile-SD are compared to the simulation results in which the validity of the proposed model is confirmed. Moreover, the output of AACPT is exploited to formulate a new equation which calculates the optimal $\lambda_{\rm max}$ from a given $\beta$ in order to conserve the standard interface of TCP. This equation increases the scalability of Agile-SD and improves its total performance.
\end{abstract}


\IEEEpeerreviewmaketitle

\section*{Introduction}
\IEEEPARstart{O}{ne} of the most predominant protocols of the Internet is the Transmission Control Protocol (TCP), which provides a high level of reliability on end-to-end connections. It regulates the transmission rate between the two ends of a connection based on the changes of the underlying network. In other words, it estimates the condition of the connection and adjusts its congestion window (\textit{cwnd}) accordingly.

In order to enhance the total performance of TCP over high-speed networks, many Congestion Control Algorithms (CCAs) have been proposed in the literature such as Scalable TCP \cite{Kelly2003}, HS-TCP \cite{Floyd2003}, H-TCP \cite{Leith2004}, BIC \cite{xu2004}, TCP Africa \cite{King2005}, TCP Compound \cite{Tan2006}, Fusion \cite{Kaneko2007}, YeAH \cite{Baiocchi2007}, TCP illinois \cite{Liu2008}, Cubic, \cite{Ha2008} and HCC \cite{xu2011}. The main three approaches, which are employed by these CCAs, are either loss-based, delay-based or loss-delay-based approach \cite{alrshah2014}.

The delay-based approach relies on the variation of delay resulted by big buffers and/or long RTTs, which are presented in high-BDP networks. However, the low-BDP networks employ small buffer sizes and generate very short RTTs resulting in a very trivial delay variation, which makes it worthless to use a delay-based approach. These characteristics of such networks allow TCP to rely only on packet losses since it is the only indicator to congestion. In turn, this behavior makes TCP very sensitive to packet losses, which negatively affects its total performance.

One of the latest versions of loss-based TCP CCA is Agile-SD \cite{alrshah2015}, which has been proposed with a view to mitigating the sensitivity to packet losses. It proposes Agility Factor ($\lambda$), which is used to shorten the epoch time (the epoch is the time needed by a CCA to increase its $cwnd$ from the time of reduction to the time of attaining the maximum utilization of the link), as shown in Fig.\ref{fig:AGILE}. The unique Agility Factor Mechanism (AFM) of Agile-SD has been evaluated by extensive simulation experiments, which confirmed that Agile-SD has the ability to reduce the sensitivity to packet losses and to improve the performance of TCP to a reasonable extent, especially over low-BDP networks.

In the literature, many mathematical models have been proposed in order to study and evaluate the performance of TCP. The authors of \cite{padhye1998,padhye2000} proposed a clear model to calculate the steady-state throughput of TCP as a function of RTT and loss rate. Also, The authors of \cite{mathis1997} evaluated the throughput of TCP using periodic loss based model. The authors of \cite{misra1999} used the Markov chains and the stationary distribution to predict the behavior of TCP dealing with RED-based routers. The authors of \cite{Shorten2007} used the random matrix model to evaluate the performance of multiple AIMD TCP flows via drop-tail queue management system. The authors of \cite{bao2010} derived the stationary distribution of Markov chains to calculate the steady-state throughput of TCP Cubic.

As aforementioned, the models in \cite{padhye1998,padhye2000,mathis1997,misra1999,Shorten2007,bao2010} calculate the steady-state throughput of TCP as a function of RTT and loss rate, while they did not take the buffer size into account. In order to calculate the throughput of TCP as a function of buffer, RTT and loss rate, we propose a novel mathematical model to calculate the average throughput of both NewReno and Agile-SD over high-speed networks. The main contributions of this model are: First, to validate the simulation results of Agile-SD by comparing it to the numerical results of this model and to the results of NewReno as a benchmark. Second, to study the impact of $\lambda_{\rm max}$ parameter on the throughput and epoch time, where the epoch time is a period of time confined between two consecutive losses. Third, to formulate an equation for automating the configuration of optimal $\lambda_{\rm max}$ based on the given system parameter $(\beta)$ in order to increase the scalability of Agile-SD.

\section{System Model for AFM-Based Agile-SD TCP}
Consider a link between two computers and suppose that the link speed is \emph{C} (in Kbps) and the source has a large file to send to the destination. Assume that the Packet Size is $\theta$ (in Kbits). Also, assume that the RTT (in seconds) is constant, which is a common assumption in loss-based TCP mathematical models \cite{padhye1998, padhye2000, bao2010}. Since the Bandwidth-Delay Product (BDP) of a link is equal to the multiplication of \emph{C} by RTT, consequently, the maximum congestion window size $W$ is calculated as Equation \eqref{eq1} \cite{huh2006},
\begin{align}
	W = \dfrac{\rm BDP}{\theta} 				\label{eq1}.
\end{align}

As for the buffer size, we necessarily need to understand its impact on the behavior of TCP to know how it could affect this model. Based on Equation \eqref{eq1}, suppose that we have a scenario in which the maximum window size is 100 packets, which allows 100 packets, maximum, to be in-flight. Then, let us say that the buffer size used in this link is 20 packets, which allows the maximum packets in-flight to be 120 packets. Accordingly, we can conclude that the buffer size is playing a role in extending the capacity of the link, thus, Equation \eqref{eq1} can be reformulated as below,
\begin{align}
	W \approx \dfrac{\rm BDP}{\theta} + b, 							\label{eq2}
\end{align}
where $b$ is a constant indicates to the buffer size in packets.

\subsection{Congestion Control of Agile-SD}
\begin{figure}[h!]
	\centering
	\includegraphics[width=0.95\linewidth]{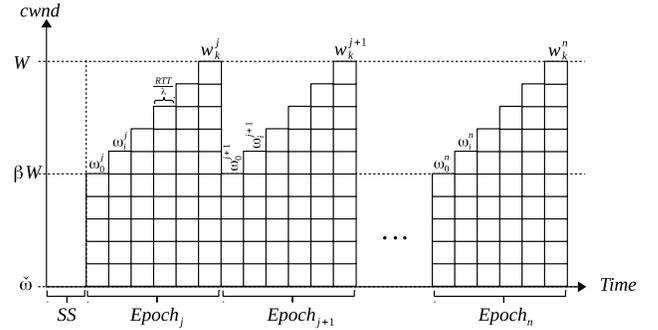} 
	\caption{\textit{cwnd} evolution of Agile-SD TCP.}
	\label{fig:AGILE}
\end{figure}

At the congestion avoidance stage, Agile-SD increases its congestion window $w$ by a small fraction after every reception of acknowledgment, similarly like the standard NewReno. However, NewReno calculates this increase as 1 over $w$, which gives:
\begin{align}
w_i^j = w_{i-1}^j + \dfrac{1}{w_{i-1}^j},				\label{eq3}
\end{align}
while in Agile-SD this increase is calculated as $\lambda$ over $w$, which gives:
\begin{align}
w_i^j = w_{i-1}^j + \dfrac{\lambda_i^j}{w_{i-1}^j},				\label{eq4}
\end{align}
where the cycle index $i$ $\in \{1, 2, \ldots, k\}$, $k$ is the amount of degradation in congestion window after loss, $k = W(1 - \beta)$, $\beta$ is the multiplicative decrease factor of TCP, the epoch index $j$ $\in \{1, 2, \ldots, n\}$ and $n$ is the number of epochs needed for TCP connection to finish its data transmission, where $n = \frac{\rm TT}{k}$ and TT is the total transmission time, as shown in Fig.\ref{fig:AGILE}. As for $\lambda^j_i$, it represents the agility factor of Agile-SD at the cycle $i$ in the epoch $j$, which is calculated as in Equation \eqref{eq8} below,
\begin{align}
	\lambda^{j}_{i} = \max\left(\lambda_{\rm max}\times \left(\dfrac{w^{j-1}_{k} - w^{j}_{i-1}}{w^{j-1}_{k} - w^{j}_{0}}\right),\lambda_{\rm min}\right)	\label{eq8},
\end{align}
where $\lambda_{\rm min}$ and $\lambda_{\rm max}$ are system parameters, $\lambda_{\rm min} = 1$, $\lambda_{\rm max} > 1$, \mbox{$\lambda^j_i \in [\lambda_{\rm min}, \lambda_{\rm max}]$}, $w_{0}^j$ is the congestion window on the time of reduction of the previous epoch, where \mbox{$w_0^j = \max(\beta w_k^{j-1},\check{w})$}, $w_k^{j-1}$ is the congestion window just before the time of reduction of the previous epoch and $\check{w}$ is a system parameter representing the minimum allowed congestion window. As for max() function, it is used to always guarantee that $\lambda^{j}_{i} \geq \lambda_{\rm min}$.

From Fig.\ref{fig:AGILE}, it is very clear that the epoch is a period of time confined between two consecutive losses \cite{bao2010}. Each epoch $j$ contains a number of sequent cycles and every cycle $i$ is a sub-period of time confined between two consecutive increases or between a consecutive degradation and increase. As known, the standard TCP variants are RTT-dependent; in which TCP needs to receive all Acks of the previous \textit{cwnd} to increase their \textit{cwnd} by one. Consequently, these TCP variants consume a complete RTT per cycle in order to achieve that increase. Since the operating systems allow their \textit{cwnd} to be only an integer number, we use the flooring function during the assignment of a new value to the \textit{cwnd} of the operating system. Thus, the Transmission Rate (Tr) of data at cycle $i$ in epoch $j$ is,
\begin{align}
{\rm Tr}_i^j \text{(Packets/s)}	 = \dfrac{ \lfloor w_i^j  \rfloor}{\rm RTT} 		\label{eq5}.
\end{align}

\begin{figure} [h!]
	\begin{center}
		\subfigure [Standard TCP]{
			\includegraphics[width=0.45\linewidth]{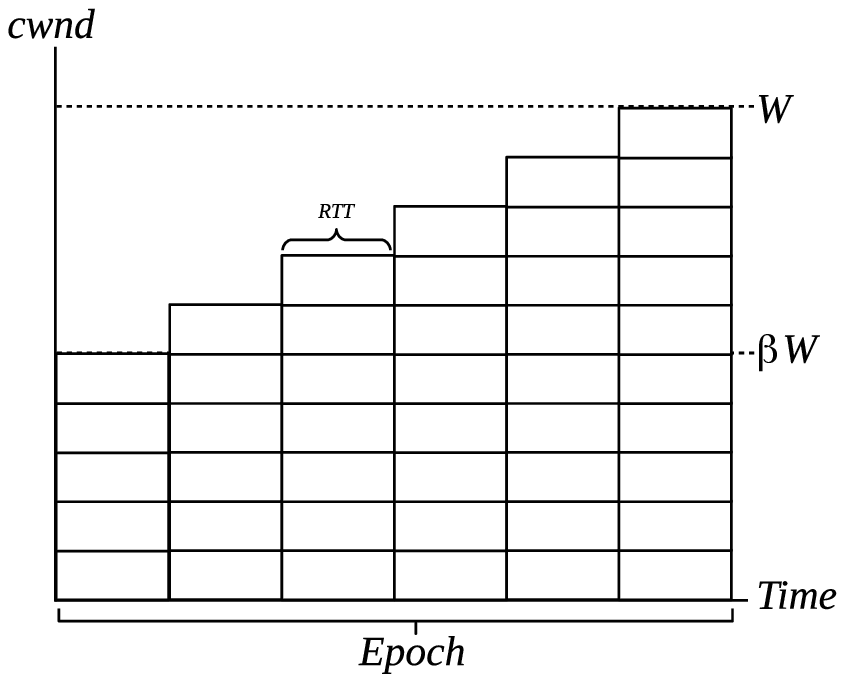}	
			\label{fig:epochsa}
		}
		\subfigure [Agile-SD TCP]{
			\includegraphics[width=0.34\linewidth]{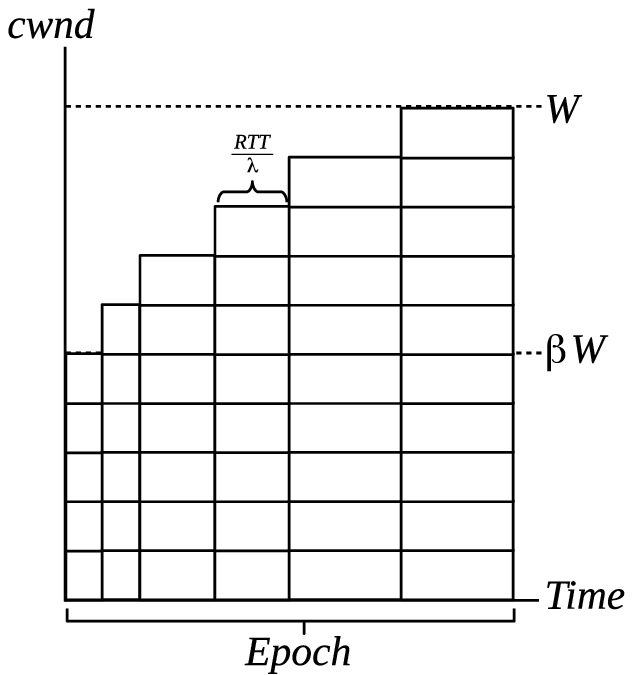}	
			\label{fig:epochsb}
		}
	\end{center}
	\caption{The evolution of cwnd.}
	\label{fig:epochs}
\end{figure}

Differently, Agile-SD is an RTT-independent, which consumes only $\frac{1}{\lambda} {\rm RTT}$ to increase its \textit{cwnd} by one. Thus, Agile-SD consumes shorter time than the time needed by the RTT-dependent TCP variants to reach the maximum \textit{cwnd}. In other words, Agile-SD shows shorter epoch time than RTT-dependent TCPs, as shown in Fig.\ref{fig:epochsa} and Fig.\ref{fig:epochsb}.

In fact, Agile-SD is still subject to the same transmission rate of standard TCP as Equation \eqref{eq5}, however, Agile-SD sends only $\frac{w}{\lambda}$ packets in every cycle, where the cycle duration time is only $\frac{\rm RTT}{\lambda}$ seconds. This behavior does not increase the transmission rate per cycle, but it controls the period of transmission time given to every cycle. Thus, it shortens the time given for the cycles with low rates and lengthens the time given to the cycles with high rates, which results in shorter epochs (as in Fig.\ref{fig:epochsb}) than the RTT-dependent TCP variants (as in Fig.\ref{fig:epochsa}).

Let us now calculate the transmission rate per epoch, which is equal to the total data sent, in an epoch, over its duration time. Since the data sent at a cycle $i$ is equal to $\frac{\lfloor w_i\rfloor}{\lambda_i}$ and the duration time of a cycle $i$ is equal to $\frac{\rm RTT}{\lambda_i}$, thus, the Epoch Average Transmission Rate (EATr$^j$) is equal to the summation of data sent over all cycles in epoch $j$ divided by the duration time of the epoch cycles, as in Equation \eqref{eq6} below,
\begin{align}
	{\rm EATr}^j({\rm Kbps}) = \theta \times \dfrac{\sum\limits_{i=0}^{k} \dfrac{\lfloor w_i^j \rfloor}{\lambda_i^j}}{\sum\limits_{i=0}^{k} \dfrac{\rm RTT}{\lambda_i^j}}, \forall j \in \{1,2,\ldots, n\}							\label{eq6},
\end{align}
where $n = \frac{\rm TT}{k}$, TT is the total transmission time, $k$ is the number of cycles in epoch $j$, \mbox{$k = W(1 - \beta)$}, $\theta$ is the packet size, $w_{0}^j$ is the congestion window at the time of reduction of the previous epoch where \mbox{$w_0^j = \max(\beta w_k^{j-1},\check{w})$}, $\lambda_0^j = \lambda_{\rm max}$, and $\beta$ is the multiplicative decrease factor of TCP.

However, calculating the transmission rate per epoch is not a final target, it is only to be used for calculating the Total Average Transmission Rate (TATr) of a connection. Fundamentally, any connection passes through an $n$ epochs, where every epoch $j$ sends an amount of data equal to $\sum_{i=0}^{k} ({\lfloor w_i^j \rfloor}/{\lambda_i^j})$ over a duration time equal to $\sum_{i=0}^{k} ({\rm RTT}/{\lambda_i^j})$. Thus, the TATr of a connection can be calculated as the summation of total data sent through the connection over the total duration of its transmission time, as in Equation \eqref{eq7} below,
\begin{align}
	&{\rm TATr(Kbps)} = \theta \times \dfrac{\sum\limits_{j=1}^{n} \sum\limits_{i=0}^{k} \dfrac{\lfloor w_i^j \rfloor}{\lambda_i^j}}{\sum\limits_{j=1}^{n} \sum\limits_{i=0}^{k} \dfrac{\rm RTT}{\lambda_i^j}}, w^{j}_{i} \leq W.		\label{eq7}
\end{align}

\subsection{Congestion Loss and Random Packet Loss}
Assume that packet losses are either \textit{congestion loss} or \textit{random packet loss}. Congestion loss happens when the transmission rate attains the maximum capacity $C$ of the bottleneck link or the maximum window size $W$, and it also happens when the buffers are overflowed. As for random packet loss, it can be caused by collision, interference and/or fading, where the random packet loss is subject to the Poisson distribution with rate $\emph{R}$ \cite{bao2010, hassayoun2010}. Thus, the probability density function $P_x$ for a given congestion window $w$ is:
\begin{align}
P_x(w) = 
\left\{\begin{array}{ll}
\dfrac{(Rw)^x e^{-Rw}}{x!} & ,x = 0, 1, 2, ...\\ \\
0 & ,\text{otherwise}
\end{array}\right.
\end{align}
where $P_x(w)$ is the probability of occurrence of $x$ packet losses in congestion window $w$.

\subsection{Markov Chain Formulation}
Let $\{w_{1}, w_{2}, \ldots, w_{N}\}$ denote the range of congestion window size, which represents a system with an $N$ states, where $w_1=\check{w}$ and $w_N=W$, which results in,
\begin{align} \label{nxn}
	N = W - \check{w} + 1,
\end{align}
where $W$ and $\check{w}$ are the maximum and minimum allowed congestion window, respectively. 

Let $T$ denote the transition probability matrix of the Markov chains for the system of $N$ states, given as,
\begin{align}
	T = 
	\begin{bmatrix}
		_{v[1,1]}   & _{v[1,2]}   & \dots 	& _{v[1,N-1]} 	& _{v[1,N]} 	\\
		_{v[2,1]}   & _{v[2,2]}   & \dots 	& _{v[2,N-1]} 	& _{v[2,N]} 	\\
		\vdots      & \vdots      & \ddots  & \vdots 		& \vdots 		\\
		_{v[N-1,1]} & _{v[N-1,2]} & \dots 	& _{v[N-1,N-1]}	& _{v[N-1,N]}	\\
		_{v[N,1]}   & _{v[N,2]}   & \dots	& _{v[N,N-1]} 	& _{v[N,N]} 	\\
	\end{bmatrix},
\end{align}
where $v[i, j]$ represents the transition probability of the system to move from the $i^{th}$ state to the $j^{th}$ state, where \mbox{$i, j \in \{1, 2, \ldots, N\}$}. If the \textit{cwnd} size is in the $i^{th}$ state, it is denoted by $w_i$. This finite set of \textit{cwnd} sizes corresponds to the Markov Chain with $N$ states. 

\begin{figure}[!h]
	\centering
	\includegraphics[width=0.85\linewidth]{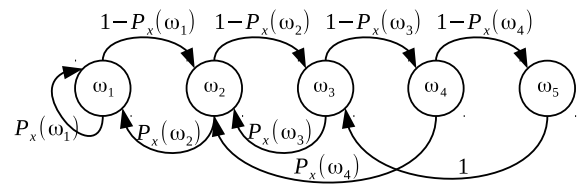}
	\caption{State transition diagram for Markov chain example of $N=5$, $\beta=0.75$, $w_i \in \{2,3,4,5,6\}$ and $i \in \{1,2,\ldots,N\}$.}
	\label{fig:tdiagram}
\end{figure}

Let us consider an example of a system with $W=6$ and $\check{w}=2$. Based on Equation \eqref{nxn}, the system has only 5 states ($N = 5$), where the finite set of the system states is $w_i \in \{2,3,4,5,6\}$ where $i \in \{1,2,\ldots,N\}$. Also, let us suppose that $\beta = 0.75$. Thus, the transition diagram of the system can be drawn as shown in Fig.\ref{fig:tdiagram}, where the transition probabilities are calculated based on the probability distribution shown in Equation \eqref{eq11} \cite{jansang2009,jansang2013} below:
\begin{align}
v[i, j] = 
\left\{\begin{array}{ll}
P_x(w) 	& , j = \lfloor\beta \times i\rfloor\\
& , i = [1, N-1]\\ \\
1-P_x(w)& , j = i+1\\
& , i = [1, N-1]\\ \\
1		& , j = \lfloor\beta \times i\rfloor\\ 
& , i = N\\ \\
0		& , \text{otherwise},
\end{array}\right. \label{eq11}
\end{align}
where $w$ is the congestion window, $v[i, j]$ is the transition probability of the system to move from the $i^{th}$ state to the $j^{th}$ state, and \mbox{$i, j \in \{1, 2, \ldots, N\}$}.

\begin{figure} [!h]
	\centering
	\includegraphics[width=0.85\linewidth]{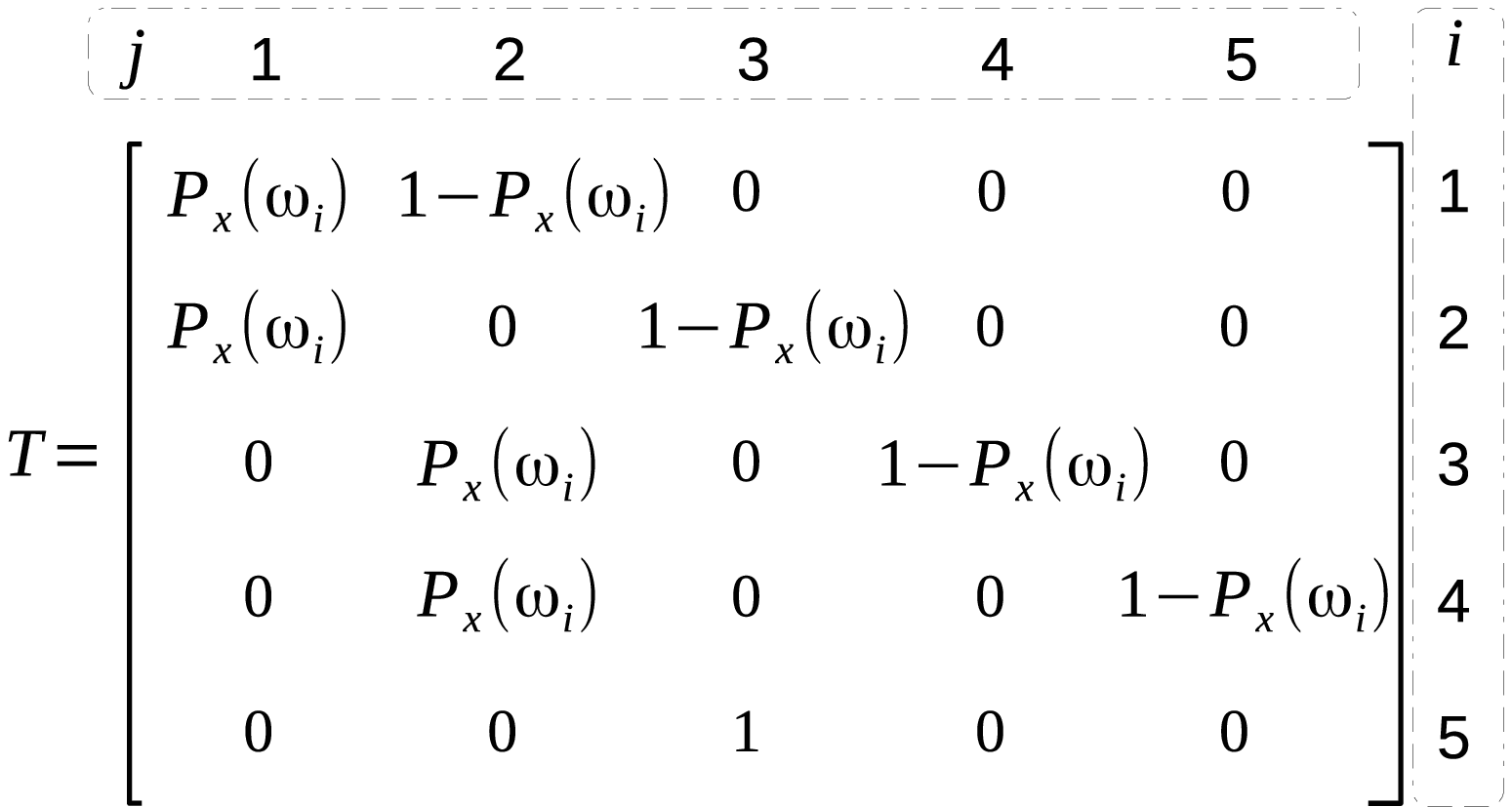}
	\caption{State transition matrix for Markov chain example of $N=5$, $\beta=0.75$, $w_i \in \{2,3,4,5,6\}$ and $i \in \{1,2,\ldots,N\}$.}
	\label{fig:tmatrix}
\end{figure}

Let us now put the probabilities of all represented transitions shown in Fig.\ref{fig:tdiagram} into their relevant places in the transition matrix $T$. Then, let us fill the unrepresented transitions by zeros. Thus, the resulted transition matrix $T$ will be as shown in Fig.\ref{fig:tmatrix}. 

Let $v_i$ be the $i^{th}$ row vector in $T$, which represents the probabilities for the system to move from the $i^{th}$ state to all possible \emph{N} states of the system,
\begin{align}
	v_i = \Big[v[i,1], v[i,2], \ldots, v[i,N]\Big],
\end{align} 
where every element represents the probability to move from the $i^{th}$ state to one $j^{th}$ state.

From Fig.\ref{fig:tmatrix}, we can see that the $v_5$ is a special case, in which the system will transit from $w_5$ to $w_3$ due to a random loss with $P(w_5)$ probability or a congestion loss with \mbox{$(1-P(w_5))$}. Thus, the total probability will be \mbox{$P(w_5) + 1-P(w_5) = 1$}.
	
In order to validate the distribution of these probabilities in the matrix $T$, the summation of the probabilities in every row vector must always be equal to one.
\begin{align}
	\sum_{j=1}^{N}v{[i,j]} = 1, \forall i \in \{1, 2, \ldots, N\}.
\end{align}

Let $v^{(t)}$ be the probability distribution of the system states at the transition $t$, where $v^{(t)}$ is equal to the product of the previous probability distribution of the system states at the transition $(t-1)$ and the matrix $T$, as follows,
\begin{align}
	\label{eqvi1}
	v^{(t)} = v^{(t-1)} \times T, \quad \forall t \in \{1, 2, \ldots, \infty\},
\end{align}
where $v^{(0)}$ is the initial state, which is the transition probability distribution of system states at the time zero. 

Since this model is for the average throughput, we set the initial \textit{cwnd} of the system to $\lceil\beta W\rceil$, which represents a system in the state of loss degradation at the congestion avoidance stage. Thus, the initial state row vector of the Markov chain is denoted by $v^{(0)}$, in which the probabilities are distributed as below:
\begin{align}
v^{(0)}[j] = 
\left\{\begin{array}{ll}
1 	& , j = \lceil\beta N\rceil\\ \\
0	& , \text{otherwise}
\end{array}\right. , \forall j  \in \{1,2,\ldots,N\}, \label{eq111}
\end{align}
where $j$ is the item index in $v^{(0)}$.

In order to calculate the Average Throughput (ATh) of the system while taking the congestion and random losses into account, we employ Equation \eqref{eqvi1} into Equation \eqref{eq7} to count all data packets received at the destination over the given connection. Since the source sends $w^t$ data at transition $t$ and the destination receives only $(v^{(t-1)} \times T) \times S^{'}$ data at the same transition, therefor, the ATh can be calculated as in Equation \eqref{ATh} below,
\begin{align}
	{\rm ATh(Kbps)} = \theta \times \dfrac{\sum\limits_{t=1}^{I} \dfrac{(v^{(t-1)} \times T) \times S^{'}}{\lambda^{(t)}}}
	{\sum\limits_{t=1}^{I} \dfrac{\rm RTT}{\lambda^{(t)}}}, \; {\rm ATh} \leq \frac{W}{\rm RTT} \label{ATh},
\end{align}
where $I$ is the number of iterations needed for the system to reach the steady-state, $S^{'}$ is the transpose of the sample space vector $S$, where $S = [w_{1}, w_{2}, \ldots,  w_{N}]$, and $\lambda^{(t)}$ is the agility factor at $\forall t \in \{1, 2, \ldots, \infty\}$, as,
\begin{align}\label{lambda}
\lambda^{(t)} = \max\left(\lambda_{\rm max}\times\left(\frac{W - \Big((v^{(t-1)} \times T) \times S^{'}\Big)}{W - (\beta W)}\right), \lambda_{\rm min}\right),
\end{align}
where $\lambda_{\rm min}$ and $\lambda_{\rm max}$ are system parameters, $\lambda_{\rm min} = 1$, $\lambda_{\rm max} > 1$, \mbox{$\lambda^{(t)} \in [\lambda_{\rm min}, \lambda_{\rm max}]$}.

Indeed, another Markovian model to account for the changing agility factor ($\lambda$) is needed to be integrated with Equation \eqref{ATh} to find the stationary distribution of the system states, by calculating the eigenvector of $T$ with eigenvalue 1. However, the integration of two Markovian models will increase the complexity of the main model and may hinder the process of understanding the model. In order for us to avoid this complexity and since $\lambda^{(t)}$ is easy to be determined at every transition $t$ in the main Markov chain of this model, we account for $\lambda^{(t)}$ using Equation \eqref{lambda} at every cycle in Equation \eqref{ATh}. Based on observation, we found that the system starts entering to the steady state after about 4 thousand iterations, however, 10 thousand iterations are used in order to increase the precision of results.

\section{Performance Evaluation}
\begin{figure}[!h]
\centering
\includegraphics[width=0.85\linewidth]{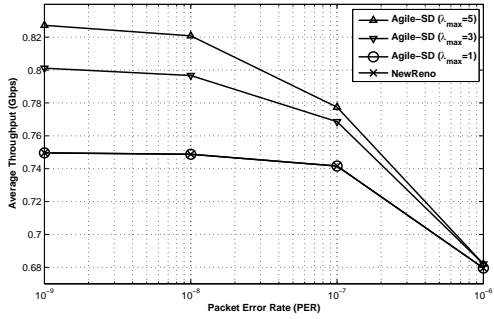}
\caption{Simulation-based comparison between Agile-SD with different $\lambda_{\rm max}$ and NewReno, under different PERs.}
\label{fig:lambda}
\end{figure}

In this section, we compare the numerical results for both NewReno and Agile-SD to the simulation results in order to confirm the validity of the proposed model. Indeed, our mathematical model is able to represent both NewReno and Agile-SD, where the numerical results of NewReno can be obtained using this model with $\lambda_{\rm max}=1$. Using simulation, we confirmed that Agile-SD would exactly perform like NewReno if $\lambda_{\rm max}$ was set to 1, as shown in Fig.\ref{fig:lambda}.

In order to obtain the numerical results, we use Octave version 4.0, which is a major free alternative software to MATLAB. More specifically, we run our model with both $\lambda_{\rm min}$ and $\lambda_{\rm max}$ equal to 1 to obtain the results of NewReno. Thereafter, we run it again with $\lambda_{\rm min}$ and $\lambda_{\rm max}$ equal to 1 and 5, respectively, to obtain the results of Agile-SD. For better understanding, Table \ref{params1} shows the setting of the model.

\begin{table}[!h]
	\centering
	\caption{The Setting of Model Parameters.}
	\label{params1} 
	\begin{tabular}{p{4.3cm}|ll}
		\hline
		\multicolumn{1}{c}{Parameter} 			& \multicolumn{2}{c}{Value(s)}            						\\ \hline 
		\multicolumn{3}{l}{} \vspace{-0.2cm}																	\\ 
		Buffer size ($b$)              			& \multicolumn{2}{l}{From 4 to 128 packets} 					\\
		Loss rate ($R$)                			& \multicolumn{2}{l}{From $10^{-8}$ to $10^{-3}$}				\\
		Packetsize ($\theta$)          			& \multicolumn{2}{l}{1000 bytes}                   				\\
		Link capacity ($C$)           			& \multicolumn{2}{l}{1Gbps}                    					\\
		2-way Link delay (RTT)       			& \multicolumn{2}{l}{10ms}                    					\\
		Minimum allowed window ($\check{w}$)    & \multicolumn{2}{l}{2}                    						\\
		Maximum allowed window ($W$)  		    & \multicolumn{2}{l}{$(C \times {\rm RTT} / \theta) + b$} 			\\
		System states ($N$)			  		    & \multicolumn{2}{l}{$(W - \check{w}) + 1$}  					\\
		Sample space of the system ($S$)	    & \multicolumn{2}{l}{$\{\check{w}, \check{w}+1, \ldots, W\}$}	\\		
		Iterations ($I$)        	   			& \multicolumn{2}{l}{10,000 state transitions}      			\\ 
		Congestion control algorithms  			& \multicolumn{1}{c|}{Agile-SD} & \multicolumn{1}{c}{NewReno}   \\ 
		\cline{2-3} 							& \multicolumn{2}{}{} 			\vspace{-0.25cm}				\\
		Multiplicative decrease factor $(\beta)$& \multicolumn{1}{c|}{0.5}      & \multicolumn{1}{c}{0.5}       \\
		Minimum $\lambda$ $(\lambda_{\rm min})$	& \multicolumn{1}{c|}{1}        & \multicolumn{1}{c}{1}         \\
		Maximum $\lambda$ $(\lambda_{\rm max})$	& \multicolumn{1}{c|}{5}        & \multicolumn{1}{c}{1}         \\
		\multicolumn{3}{l}{} \vspace{-0.2cm}																	\\ \hline
	\end{tabular} 
\end{table}

Further, we present the impact of some parameters, such as buffer size and $\lambda_{\rm max}$, on the epoch time and average throughput of Agile-SD. At the end, an Automated Algorithm Configuration and Parameter Tuning (AACPT) \cite{Hamadi2013} process is exploited to automate the calculation of the optimal $\lambda_{\rm max}$ for a given multiplicative decrease factor $\beta$, instead of configuring it manually by the system administrators as a preset parameter.

\subsection{Model Validation via Simulation}
\begin{figure}[!h]
	\centering
	\includegraphics[width=0.85\linewidth]{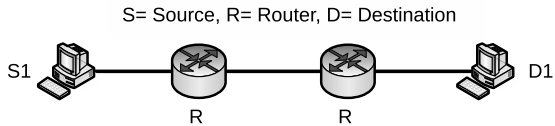}
	\caption{The network topology used for simulation}
	\label{fig:topology}
\end{figure}

For simplicity, a node-to-node topology, as shown in Fig.\ref{fig:topology}, is used to validate the results of this mathematical model using NS2 simulator version 2.35. The sender-node (S1) sends an FTP data to the receiver-node (D1) for 100 seconds over a full-duplex TCP connection, in which the bandwidth is 1 Gbps, the two-way propagation delay is 10 milliseconds, and the buffer size in each node is varied from 4 to 128 packets subject to Drop-tail, while the PER is varied from $10^{-8}$ to $10^{-3}$. Furthermore, $\beta$ and $\lambda_{\rm max}$ are set to 0.5 and 5, respectively. In fact, these setting (as shown in Table \ref{params}) are used to mimic the worst short-distance network configurations, where the buffer sizes are very small and the PERs are very high.

\begin{table}[!h]
	\caption{The Setting of Simulation Parameters.}
	\begin{center}
		\begin{tabular}{p{4.3cm}|l}
			\hline
			\multicolumn{1}{c}{Parameter}	&	\multicolumn{1}{c}{Value(s)}\\ \hline
			\multicolumn{2}{l}{} \vspace{-0.2cm}							\\ 
			CCAs							&	Agile-SD, NewReno			\\ 
			$\beta$ 						&	0.5							\\ 
			$\lambda_{\rm min},\lambda_{\rm max}$ 	&	1, 5 only for Agile-SD		\\ 
			Buffer size						&	From 4 to 128 packets   	\\ 
			Packet size						&	1000 bytes					\\ 
			Link capacity					&	1Gbps						\\ 
			2-way Link delay				&	10ms 						\\ 
			Loss rate (PER)					&   From $10^{-8}$ to $10^{-3}$	\\ 
			Queuing Algo	 				&	Drop-Tail					\\ 	
			Traffic type					&	FTP							\\ 
			SACK, FACK						&	Disabled					\\ 
			Simulation time					&	100 seconds					\\ 
			Simulator version				&	NS-2 ver2.35				\\ 
			\multicolumn{2}{l}{} \vspace{-0.2cm}							\\ \hline
		\end{tabular}
		\label{params}
	\end{center}
\end{table}

\begin{figure}[h!]
	\centering
	\includegraphics[width=0.85\linewidth]{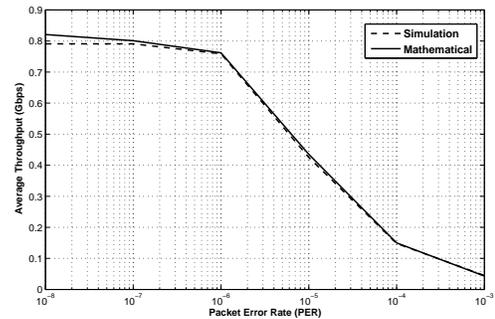}
	\caption{Agile-SD normalized average throughput under different PERs.}
	\label{fig:LR}
\end{figure}

In order to obtain the simulation results shown in Fig.\ref{fig:LR}, where the buffer size is fixed to $4$ packets, we run the simulation for 10 times to calculate the average throughput for every PER. As for the simulation results shown in Fig.\ref{fig:Comparison}, where the PER is fixed to $10^{-8}$, we also run the simulation for 10 time to calculate the average throughput for every buffer size. Additionally, the numerical results are collected using the proposed model under the same conditions used in the simulations in order to present a fair comparison.

As it is clear in Fig.\ref{fig:LR}, the simulation and mathematical results are very close to each other. However, the mathematical curve slightly diverges from the simulation curve, due to neglecting the slow start phase in our model and also due to the additional processing overhead occur during the simulation. Nevertheless, the results in Fig.\ref{fig:LR} confirms the validity of this model supported by the results in Fig.\ref{fig:Comparison} and Fig.\ref{fig:rtt}, which compare between the throughput of Agile-SD and the standard NewReno under different conditions of buffer size and RTT, respectively.

\subsection{Average Throughput of Agile-SD}
\begin{figure}[h!]
	\centering
	\includegraphics[width=0.85\linewidth]{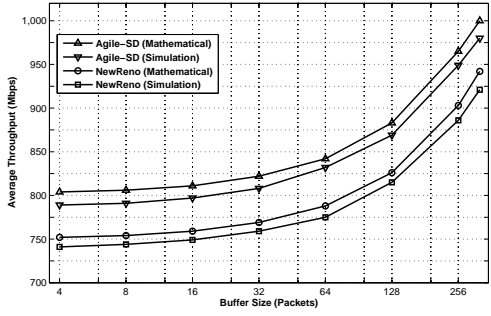}
	\caption{Normalized average throughput under different buffer sizes and $10^{-8}$PER where $\beta = 0.5$ and $\lambda_{\rm max} = 5$.}
	\label{fig:Comparison}
\end{figure}

\begin{figure} [h!]
\centering
\includegraphics[width=0.85\linewidth]{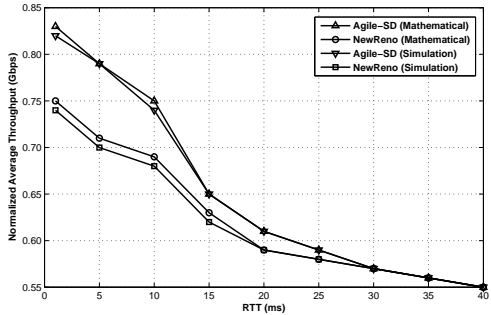}
\caption{Normalized average throughput under different RTTs and $10^{-8}$PER where $\beta = 0.5$, $\lambda_{\rm max} = 5$ and buffer size is only 4 packets.}
\label{fig:rtt}
\end{figure}

As shown in Fig.\ref{fig:Comparison}, Agile-SD overcomes the standard NewReno in all cases. Nevertheless, both algorithms show the same pattern for a proportional relationship between throughput and buffer size, which supports our assumption in Equation \eqref{eq2}. As well as, Agile-SD can perform better than the standard NewReno in most RTT cases, especially when the RTT and the used buffer size are very small, as shown in Fig.\ref{fig:rtt}. Since these network characteristics are seen in short-distance and near-zero buffer networks (such as fiber optic networks), thus, the behavior of Agile-SD would be promising for improving TCP performance if such networks are targeted.

\subsection{The Impact of Agile-SD on Epoch Time}
\begin{figure}[h!]
	\centering
	\includegraphics[width=0.85\linewidth]{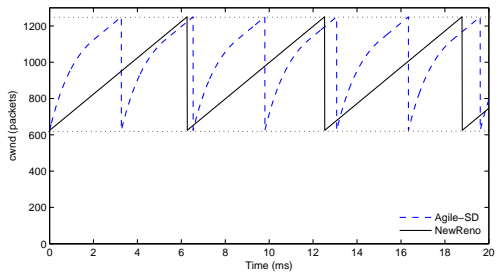}
	\caption{Agile-SD vs. NewReno epoch time.}
	\label{fig:cwnd}
\end{figure}

Due to its unique Agility Factor Mechanism (AFM), Agile-SD shows shorter epoch time than the standard NewReno, as shown in Fig.\ref{fig:cwnd}. This behavior reduces the underutilized area of bandwidth and, in turn, it improves the average throughput of TCP, as shown in Equation \eqref{ATh}, where the epoch time is represented by $\frac{\rm RTT}{\lambda}$ at the denominator. 

\begin{figure} [t!]
	\centering
	\includegraphics[width=0.95\linewidth]{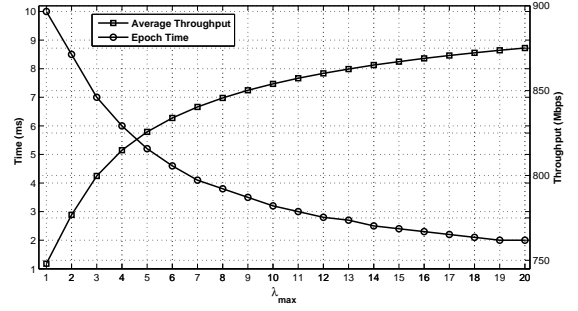}
	\caption{Impact of $\lambda_{\rm max}$ on the average throughput and epoch time.}
	\label{fig:lambda-impact}
\end{figure}

As shown in Fig.\ref{fig:lambda-impact}, it is very clear that the epoch time is inversely correlated to the value of $\lambda_{\rm max}$ parameter, while the average throughput is positively correlated to $\lambda_{\rm max}$. Thus, the greater $\lambda_{\rm max}$, the shorter epoch time and the higher average throughput, and vice versa. Hence, it can be deduced that the main player, which directly affects the performance of Agile-SD, is the parameter $\lambda_{\rm max}$. Thus, this parameter $(\lambda_{\rm max})$ must be carefully configured based on the value of $\beta$.

\section{The AACPT Process}
The designers and system users of parameterized algorithms routinely encounter a problem of how to find the optimum configurations or parameter settings to obtain the best possible results. In this section, an AACPT \cite{Hamadi2013} technique is used to find the optimal parameter settings of $\lambda_{\rm max}$ based on a set of problem instances of $\beta$. The optimum value of $\lambda_{\rm max}$ should reflect the minimum setting to obtain the maximum average throughput for a given $\beta$.

Henceforth, $\lambda_{\rm max}$ will be denoted by $\lambda'$ for the sake of simplifying the presentation. Let $\overrightarrow{\lambda'}$ donate the set of possible configurations of $\lambda'$, and let $\overrightarrow{\beta}$ denote the set of problem instances of $\beta$, where $\overrightarrow{\lambda'} = [\lambda'_1, \lambda'_2, \ldots, \lambda'_n]$ and $\overrightarrow{\beta} = [\beta_1, \beta_2, \ldots, \beta_m]$. Let $\beta\lambda'$ be the parameter configuration combination matrix with the size $[m \times n]$, which contains all possible $\beta\lambda'_{ij}$ combinations, where $\beta\lambda'_{ij} = (\beta_i, \lambda'_j)$, $\beta_i \in \overrightarrow{\beta}$, $\lambda'_j \in \overrightarrow{\lambda'}$, $i \in \{1,2,\ldots,m\}$, and $j \in \{1,2,\ldots,n\}$. Thus, each row $\overrightarrow{\beta\lambda'_i}$ in $\beta\lambda'$ merely denotes the possible $\overrightarrow{\lambda'}$ configurations for a given problem instance $\beta_i$, as shown in Equation \eqref{eq-row},
\begin{align}
	\overrightarrow{\beta\lambda'_i} = \Big[(\beta_i, \lambda'_1), (\beta_i, \lambda'_2), \ldots, (\beta_i, \lambda'_n)\Big] \label{eq-row}, \forall i \in \{1, 2, \ldots, m\}.
\end{align}

\begin{figure} [h!]
	\centering
	\includegraphics[width=0.85\linewidth]{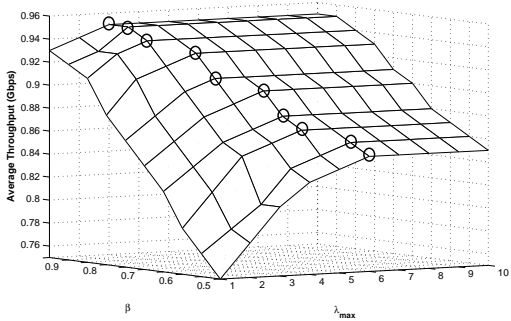}
	\caption{The normalized average throughput of Agile-SD under different configurations.}
	\label{fig:3d}
\end{figure}

Let us now calculate the Average Throughput (AT) of Agile-SD for each configuration combination $\beta\lambda'_{ij}$ using Equation \eqref{ATh}, where the result is stored into the relevant AT$_{ij}$ element in AT, which is an $[m \times n]$ matrix. Let us also track $\lambda'^i_{\rm opt}$, which is the optimal $\lambda'$ configuration in $\overrightarrow{\lambda'}$ for a given $\beta_i$. Then, let us save every $\lambda'^i_{\rm opt}$ into the relevant position in the vector of the optimum setting ($\overrightarrow{\lambda'_{\rm opt}}$), which is an $[1 \times m]$ matrix. For more understanding, Algorithm \ref{algo01} explains the process of AACPT in detail. Afterward, the AT results matrix of Agile-SD under all possible configurations is drawn as a \mbox{3-Dimensional} graph, as shown in Fig.\ref{fig:3d}, to show the complete perspective behind this AACPT. 

\begin{figure} [h!]
	\centering
	\includegraphics[width=0.85\linewidth]{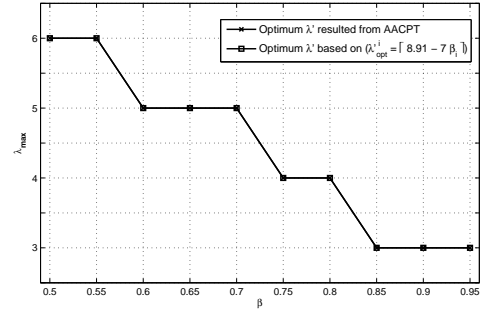}
	\caption{The relationship between $\lambda'_{\rm opt}$ and $\beta$.}
	\label{fig:APC}
\end{figure}

The circled points in Fig.\ref{fig:3d} reflect the vector of the optimal configurations $(\overrightarrow{\lambda'_{\rm opt}})$ for the given $\overrightarrow{\beta}$. For each $\beta_i$ in problem instance space $\overrightarrow{\beta}$, the optimal $\lambda'$ configuration is $\lambda'^i_{\rm opt}$ from $\overrightarrow{\lambda'_{\rm opt}}$. The \mbox{2-Dimensional} relationship between the problem instance space $\overrightarrow{\beta}$ and its optimal configuration space $\lambda'^i_{\rm opt}$, as shown in Fig.\ref{fig:APC} is directly extracted from Fig.\ref{fig:3d}.

In order to facilitate the process of formulating an equation to fit this relationship, the free on-line tool, namely \mbox{MyCurveFit} \cite{MyCurveFit}, is used to find the trend line of the optimal points. Eventually, the employment of \mbox{MyCurveFit} tool produces a simple linear equation, which is able to directly calculate $\lambda'^i_{\rm opt}$ based on a given $\beta_i$, as shown in Equation \eqref{eq-opt},
	\begin{align}
	\label{eq-opt}
	\lambda'^i_{\rm opt} =& \lceil 8.91 - 7\beta_i \rceil, \forall i \in \{1,2,\ldots,m\},
	\end{align}
where $m$ is the length of $\overrightarrow{\lambda'_{\rm opt}}$, $\beta_i \in \overrightarrow{\beta} \text{ and } \lambda'^i_{\rm opt} \in \overrightarrow{\lambda'}$. Fortunately, the result of this equation is identical to the result of the aforementioned AACPT process, as shown in Fig.\ref{fig:APC}.

\SetKwProg{Function}{Function}{ }{end}

\begin{algorithm}[h!]
	\caption{The AACPT Process.}\label{algo01}
	
	\textbf{Initialization:}\\
	\hspace{0.2cm}$\overrightarrow{\beta} \leftarrow [0.5, 0.55, 0.6, 0.65, 0.7, 0.75, 0.8, 0.85, 0.9, 0.95]$;\\ 
	\hspace{0.2cm}$\overrightarrow{\lambda'} \leftarrow [1, 2, 3, 4, 5, 6, 7, 8, 9, 10 ]$;\\
	\hspace{0.2cm}define $\beta\lambda',$ AT as $[10 \times 10]$ matrices;\\
	\hspace{0.2cm}define $\overrightarrow{\lambda'_{\rm opt}}$	as $[1 \times 10]$ matrix;\\
	\hspace{0.2cm}set $\lambda'_{\rm opt} \leftarrow 0, {\rm AT_{max}} \leftarrow 0$;\\
	
	\Function{void RunAACPT()}
	{
		\For{$i \leftarrow 1$ \KwTo $m$}
		{
			\For{$j \leftarrow 1$ \KwTo $n$}
			{
				$\beta\lambda'_{ij} \leftarrow (\beta_i, \lambda'_j)$;\\
				AT$_{ij} \leftarrow$ Agile-SD($\beta\lambda'_{ij}$);\\
				\If{$({\rm AT}_{ij} > {\rm AT}_{\rm max})$}
				{ 
					${\rm AT}_{\rm max} \leftarrow {\rm AT}_{ij}$;\\
					$\lambda'_{\rm opt} \leftarrow \lambda'_j$;\\				
				} 
			}
			${\overrightarrow{\lambda'}_{\rm opt}^i} \leftarrow {\lambda'}_{\rm opt};$
		}
		Plot3D($\overrightarrow{\beta}, \overrightarrow{\lambda'}, {\rm AT}$); $\quad \backslash\backslash$"as shown in Fig.\ref{fig:3d}"\\
		Plot2D($\overrightarrow{\beta}, \overrightarrow{\lambda'_{\rm opt}}$); $\quad\quad \backslash\backslash$"as shown in Fig.\ref{fig:APC}"
	}
\end{algorithm}

\begin{figure} [h!]
\centering
\includegraphics[width=0.85\linewidth]{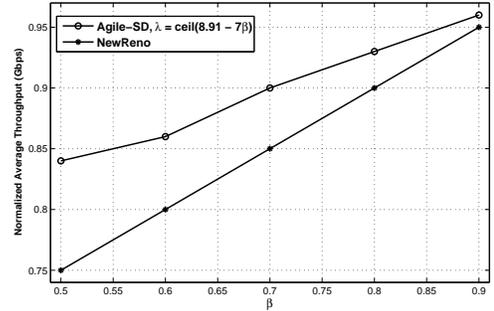}
\caption{The impact of using Equation \eqref{eq-opt} on the average throughput of Agile-SD compared to NewReno, $\beta=\{0.5 \to 0.9\}$, PER$=10^{-8}$ , and buffer size is only 4 packets.}
\label{fig:optimumlambda}
\end{figure}

It is very clear that Equation \eqref{eq-opt} has the ability to improve the average throughput of Agile-SD compared to NewReno without the need for manually configuring the $\lambda'$, as shown in Fig.\ref{fig:optimumlambda}. Thus, this equation is highly recommended to be used in Agile-SD since it can automate the calculation of the optimum value of $\lambda'$ parameter based on the preset value of $\beta$, where $\beta$ is one of the main parameters in the standard TCP interface. This automation of $\lambda'$ increases the scalability of Agile-SD and maximizes its throughput. More importantly, it helps for keeping the standard interface of TCP as it is, which facilitates the implementation of Agile-SD into the real operating systems.

\section{Conclusion}
In this paper, we propose a mathematical model to calculate the average throughput of Agile-SD and NewReno. This model is designed based on Markov chains, in which the congestion loss and packet error rate are considered. In order to validate the correctness of this model, a number of simulation experiments are carried out, in which the results are compared to the outputs of this model to validate the results presented in \cite{alrshah2015}. Besides, this model evaluates the average throughput of Agile-SD under different PERs, RTTs, and buffer sizes, in which Agile-SD overcomes NewReno even in the cases of small buffers and short RTTs. Based on this results, Agile-SD could be a promising congestion control algorithm for the short-distance networks, especially that with near-zero buffers such as fiber optic networks.

Also, the proposed model shows that the average throughput is positively correlated to the values of $\beta$ and $\lambda'$. Thus, the greater the $\beta$ and $\lambda'$, the higher the throughput and vice versa. From the other side, there must be an inverse correlation between $\beta$ and $\lambda'$ in order to keep the aggressiveness of Agile-SD balanced. In other words, whenever $\beta$ is increased, $\lambda'$ must be decreased and vice versa. For more robustness, an AACPT process is exploited to formulate an equation, which calculates the optimal configuration of $\lambda'$ based on a given $\beta$, as in Equation (\ref{eq-opt}). For future work, we plan to implement this equation in Agile-SD to make its interface compatible with the standard TCP interface.

\section*{Acknowledgment}
This work has been partially supported by the Malaysian Ministry of Education under the Fundamental Research Grant FRGS/1/2014/ICT03/UPM/01/1 for financial support.

\ifCLASSOPTIONcaptionsoff
  \newpage
\fi



\begin{IEEEbiography}[{\includegraphics[width=1in,height=1.25in,clip,keepaspectratio]{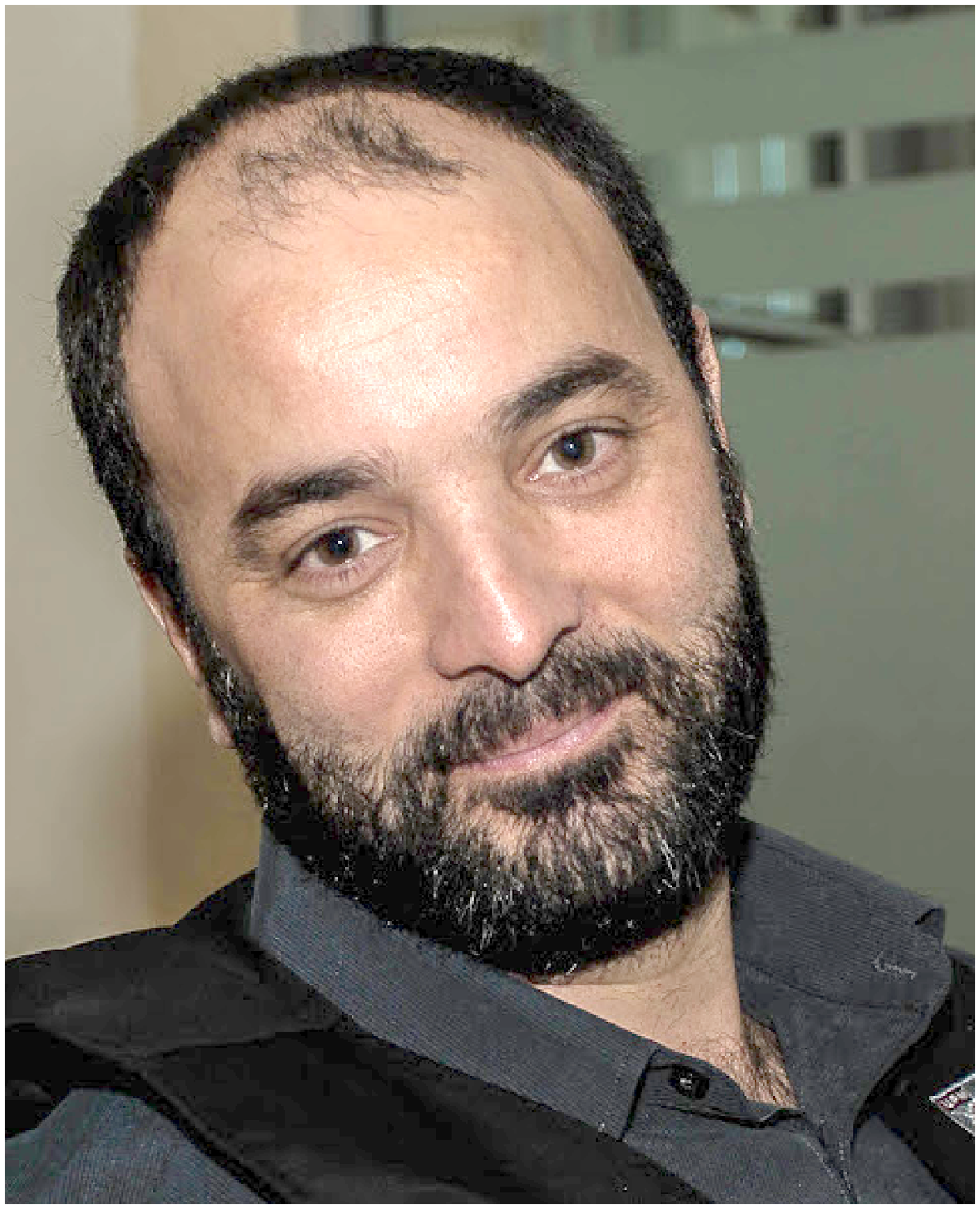}}]{Mohamed A. Alrshah}
received his BSc degree in Computer Science from Naser University - Libya, in 2000, and his MSc degree in computer networks in May 2009 from Universiti Putra Malaysia. And now he is a Ph.D candidate in the Faculty of Computer Science and Information Technology, Universiti Putra Malaysia. He has published a number of articles in high impact factor scientific journals. His research interests are in the field of high speed TCP protocols, high speed network, parallel and distributed algorithms, software defined networking, network design and management, wireless networks.
\end{IEEEbiography}

\begin{IEEEbiography}[{\includegraphics[width=1in,height=1.25in,clip,keepaspectratio]{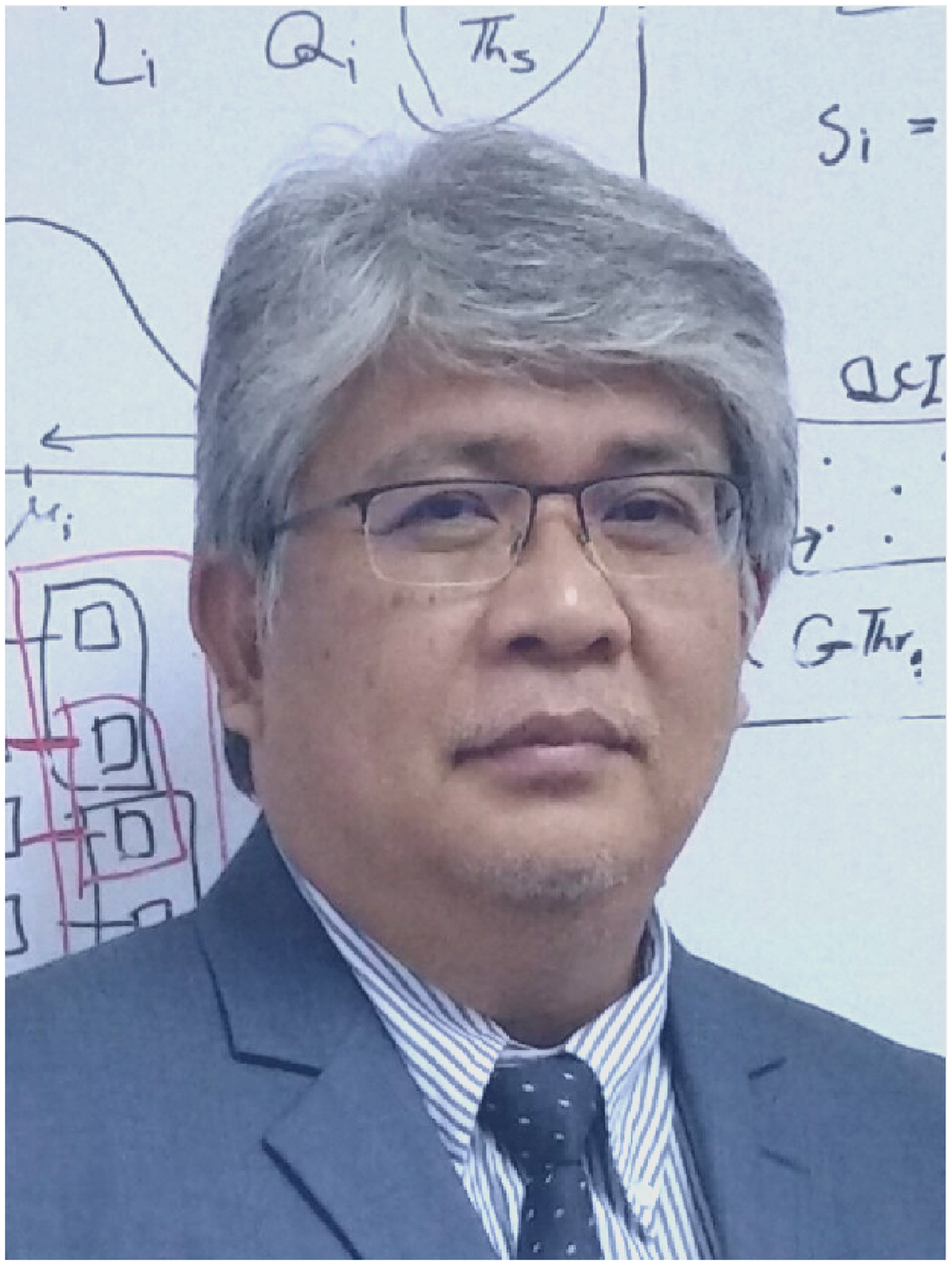}}]{Mohamed Othman}
received his PhD from the Universiti Kebangsaan Malaysia with distinction (Best PhD Thesis in 2000 awarded by Sime Darby Malaysia and Malaysian Mathematical Science Society). Now, he is a Professor in the Faculty of Computer Science and Information Technology, Universiti Putra Malaysia (UPM). He is also an associate researcher at the Lab of Computational Science and Mathematical Physics, Institute of Mathematical Research (INSPEM), UPM. He published more than 160 International journals and 230 proceeding papers. His main research interests are in the fields of high speed network, parallel and distributed algorithms, wireless network (MPDU- and MSDU-Frame aggregation, TCP Performance, MAC layer, resource management, and traffic monitoring) and scientific telegraph equation and modeling.
\end{IEEEbiography}

\begin{IEEEbiography}[{\includegraphics[width=1in,height=1.25in,clip,keepaspectratio]{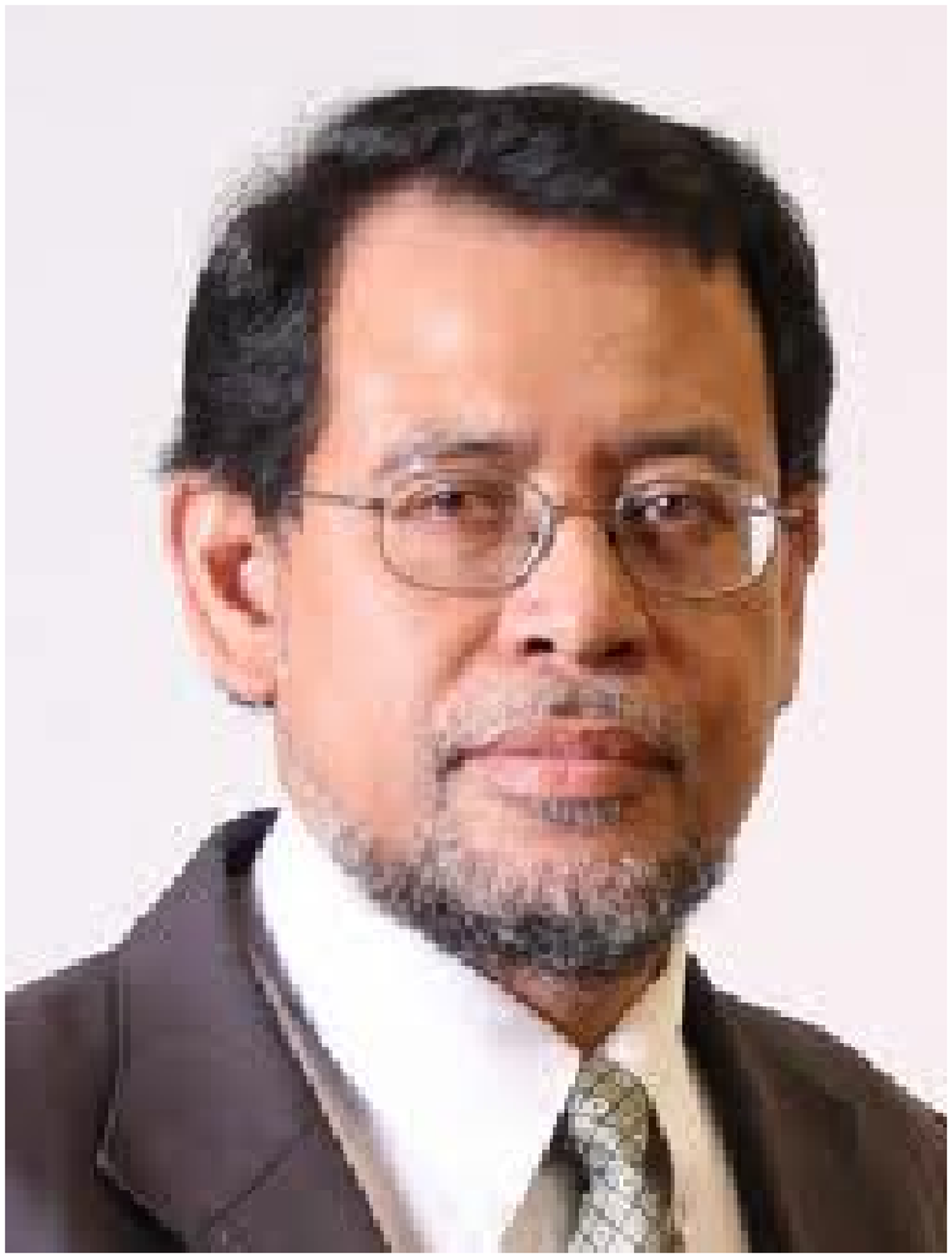}}]{Borhanuddin Mohd. Ali}
received his B.Sc (Hons.) Electrical and Electronics Engineering from Loughborough University of Technology in 1979, his M.Sc and PhD in Electromagnetic Engineering, from the University of Wales, in 1981-1985, respectively. Borhanuddin is now a full time professor in Department of Computer and Communication Engineering, Faculty of Engineering UPM. He was on secondment to MIMOS Bhd, a government research lab on ICT, heading the centre of Wireless Communications. Previously, he served as the director of the Institute of Multimedia and Software, a Centre of Excellence within the same university, 2002-2006, spent 1 year at Celcom R\&D in 1995 as visiting Scientist, and 2 years at Might in 1996-97 as a Senior Manager, charged with coming up with research and policy direction for Malaysian Telecommunication industry. In 1996 he helped to realize the formation of Teman project, and later was made the Chairman of the MyREN Research Community, a national research testbed. He is a Chartered Engineer and a member of the IET, and Senior Member of IEEE. He was the Chair of IEEE Malaysia Section 2002-2004, and previously the Chair of ComSoc Chapter, 1999-2002. His research interest spans Wireless and Broadband Communications, and Network Engineering.
\end{IEEEbiography}

\begin{IEEEbiography}[{\includegraphics[width=1in,height=1.25in,clip,keepaspectratio]{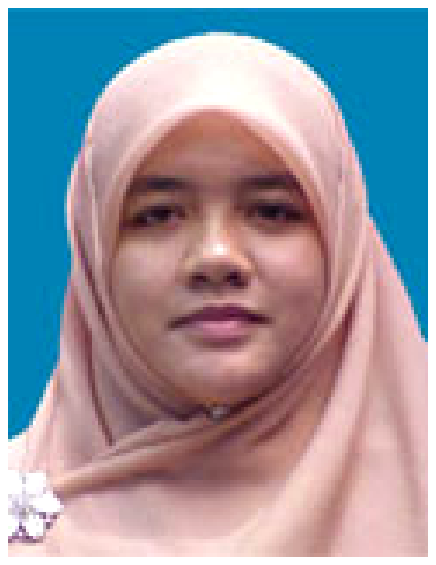}}]{Zurina Binti Mohd Hanapi}
received her B.Sc. in Computer and Electronic System Engineering, Strathclyde University in 1999, her M.Sc. in Computer and Communication Systems Engineering, Universiti Putra Malaysia in 2004 and her Ph.D.in Electrical, Electronic, and System Engineering, Universiti kebangsaan Malaysia in 2011. Now, she is working as a lecturer at Universiti Putra Malaysia. She has received an Excellence Teaching Awards in 2005, 2006 and 2012. And she has received silver medal in 2004 and bronze medal in 2012. She is a leader of some research projects and she has published many conference and journal papers. Her research interests in Routing, Wireless Sensor Network, Wireless Communication, Distributed Computing, Network Security, Crpytography, and Intelligent Systems.
\end{IEEEbiography}

\end{document}